\documentclass[preprint,amsmath,amssymb,nofootinbib,floatfix,longbibliography,natbib]{revtex4-1}
\usepackage{pslatex,graphicx,dcolumn,bm,natbib,amssymb,amsmath,color,mathtools}
\usepackage{hyperref}

\newcommand{\be}{\begin{equation}}
\newcommand{\ee}{\end{equation}}

\newcommand{\vm}{vertex model~}
\newcommand{\hh}{heterogeneity~}

\begin{document}

\title{Mechanical \hh in  tissues promotes rigidity and controls cellular invasion}

\author{Xinzhi Li, Amit Das, Dapeng Bi}
\affiliation{Department of Physics, Northeastern University, MA 02115, USA}

\begin{abstract}
We study the influence of cell-level mechanical heterogeneity in epithelial tissues using a vertex-based model. Heterogeneity in single cell stiffness is introduced as a quenched random variable in the preferred shape index($p_0$) for each cell. We uncovered a crossover scaling for the tissue shear modulus, suggesting that tissue collective rigidity is controlled by a single parameter $f_r$, which accounts for the fraction of rigid cells.  Interestingly, the rigidity onset occurs at  $f_r=0.21$, far below the  contact percolation threshold of rigid cells. 
Due to the separation of rigidity and contact percolations,  heterogeneity can enhance  tissue rigidity and gives rise to an intermediate solid state. 
The influence of heterogeneity on tumor invasion dynamics is also investigated.
There is an overall impedance of invasion as the tissue becomes more rigid.  Invasion can also occur  in the intermediate heterogeneous solid state and is characterized by significant spatial-temporal intermittency. 

\end{abstract}

\maketitle 

Heterogeneity amongst cells in a tumor has been recognized as one of the hallmarks of cancer\cite{Fidler_1978, Marusyk_2010,Weinberg_Cell_2011,f_1000_review_ITH}. This so-called intra-tumor \hh is thought to facilitate metastasis\cite{Fidler_1978, Huang_2009, Shin_2014} by allowing a cellular community the flexibility and efficiency to adapt to new environments\cite{ Ackermann_2015}. It is also largely responsible for therapeutic resistance, leading to the failure to predict the progression patterns of metastasis\cite{Fidler_1978, Marusyk_2010, Bhatia_2012, Claudi_2014}. Cellular differences within a tumor  result from an interplay  of both genetic and extrinsic influences\cite{Austin_review, Anderson_2010, Snuderl_2011, Notta_2011,Marusyk_2012}. Whereas fitness or genotypic \hh  has been studied extensively, the role of  mechanical  \hh in a tumor or cellular collective is still not well understood.  

Recent experimental evidence\cite{liver_cancer_mech_hetero,Review_The_Roles_of_Cellular_Nanomechanics_in_Cancer, Fritsch_nphys,Kaes_optical_stretcher_2017} have revealed that tumor cells exhibit a broad distribution of various biomechanical  properties. 
These include intra-tumor \hh in cell stiffnesses\cite{Plodinec_2012,fuji_afm_monolayer,Guo_2014,Fritsch_nphys,brain_tumor_afm_nanomech},
stresses,
%stresses\cite{mapping_tumor_solid_stress},
%
and cell-cell interactions\cite{Guo_2014,wirtz_pulsating_migration}. However, there is no consensus on how these heterogeneities affect the mechanical behavior at the tissue level. For example, while biophysical techniques\cite{Are_cancer_cells_really_softer_than_normal_cells} such as AFM\cite{brain_tumor_afm_nanomech} and optical stretcher\cite{Kaes_optical_stretcher_2017} show that individual cancer cells are softer than healthy cells, there is an apparent paradox with measurements at the tissue level, which show that tumors are more rigid than healthy tissues\cite{valerie_cancer_review,Sinkus_2000}. 

Similarly, from a theoretical and modeling perspective, there is a lack of understanding of how heterogeneity exhibited by single cells influence the mechanical state of the tissue. One large class of vertex-based computational models\cite{Nagai_PMB_2001, Farhadifar_CB_2007, Fletcher_BJ_2014} has been used to study mechanics of \emph{homogeneous} epithelial tissues which can exhibit a solid to fluid transition\cite{Staple_PJE_2010,bi_nphys_2015,kim_hilgenfeldt_sm_2015}. However, previous works often treat \hh as a biological noise and therefore  unable to model many of the salient features in tumors. 

In this work, we explicitly study the effect of \hh on the mechanics of the confluent cell monolayer using the \vm. 
Our results show that \hh always acts to enhance the rigidity of a tissue. The mode of enhancement by \hh depends on the spatial distribution of rigid vs. soft cells, which is directly tuned by a single universal parameter $f_r$. This results  in distinct mechanical regimes that arise from the mismatch of two percolation processes: (1) rigidity percolation of mechanical tensions in a tissue and (2) contact percolation of rigid cells. We also connect tissue rigidity to cellular migration using model for the invasion of a single metastatic cell in a heterotypic microenvironment  of the tissue.

{\bf{Model}} 
We focus on the mechanics of a 2D confluent epithelium. Similar to previous vertex model implementations, the tissue is governed by the energy function\cite{Nagai_PMB_2001,Farhadifar_CB_2007, Staple_PJE_2010, Li_Sun_Biophys,Fletcher_BJ_2014,kim_hilgenfeldt_sm_2015}
$E= \sum_{i=1}^N \left[ K_A (A_i-A_0^i)^2+ K_P (P_i-P_0^i)^2 \right]$, where
cell areas $\{A_i\}$ and perimeters $\{P_i\}$ are functions of the position of vertices $\{{\bf r}_i\}$.
$K_A$,$K_P$ are the area and perimeter elasticities. 
The term quadratic in cell area $A_i$ results from resistance to volume changes\cite{Farhadifar_CB_2007,Staple_PJE_2010,Fletcher_BJ_2014}.
Changes to cell perimeters are directly related to the deformation of the acto-myosin cortex\cite{Hufnagel_PNAS_2007, Zehnder_BJ_2015}. The term $K_P P_i^2$ corresponds to the energy cost of deforming the cortex. The linear term, $- 2 K_P P_0 P_i$, represents the effective line tension by cell-$i$  which gives rise to a `preferred perimeter' $P_0$. The value of $P_0$ can be modified by an interplay of cell-cell adhesion and cortical tension\cite{Farhadifar_CB_2007, Staple_PJE_2010, bi_nphys_2015}.
Here we assume the individual preferred cell area  $A_0$ does not vary from cell-to-cell and is set to the average area per cell (i.e. $A_0=\bar{A}$).
The tissue energy can be non-dimensionalized by choosing $K_P\bar{A}$ as the energy unit and $\sqrt{\bar{A}}$ as the length unit
\be 
\varepsilon= \sum_{i=1}^N \left[ \kappa_A (a_i-1)^2+  (p_i-p_0^i)^2 \right].
\label{eq:total_energy_non_dim}
\ee
where $a_i=A_i/\bar{A}$ and $p_i=P_i/\sqrt{\bar{A}}$ are the rescaled area and perimeter of the $i^{th}$ cell. $\kappa_A=K_A\bar{A}/K_P$ and $p^i_0=P^i_0/\sqrt{\bar{A}} $ is the preferred \emph{cell shape index}\cite{bi_nphys_2015}.

In this model, the cell stiffness is determined by the tension $\tau_m$ on cell-cell junctions (edges)\cite{Hutson_Shane_2003,Rauzi_2008, Brodland_2010, Chiou_Shraiman_2012, Ishihara_2012, Ishihara_compare_EPJE_2013, Brodland_2014}, which in turn is directly tuned by the preferred cell shape index $p_0$\cite{yan_bi_prx,Yang_PNAS_2017},
\be
\tau_{m} = (p_{i}-p_0^i) + (p_{j}-p_0^j),
\label{tension_def}
\ee
where $p_{i}$ and $p_{j}$ are the actual perimeters of cells $i,j$ adjacent to the edge $m$.
To capture the experimentally observed variations in single cell stiffnesses and in cell-cell interactions\cite{Plodinec_2012,fuji_afm_monolayer,Guo_2014,Fritsch_nphys,brain_tumor_afm_nanomech}, we introduce variations in the preferred shape indices $p_0$. The majority of this work uses a Gaussian distributed set of $p_0$'s with mean $\mu$ and standard deviation $\sigma$. 
We also performed analysis using uniform distributions with the same ($\mu,\sigma$)(SI Section I).

To initialize the simulation, Voronoi cells\cite{Bi_PRX_2016} are used to provide a set of initial vertex positions, $\{\bm r_i\}$. Then each cell is assigned a value  of $p^i_0$ drawn randomly from a Gaussian distribution with mean $\mu$ and standard deviation $\sigma$. The set of $\{p_0^i\}$ then remains as \emph{quenched variables} throughout the simulation. A combination of FIRE\cite{Bitzek_PRL_Fire_2006} and conjugate-gradient\cite{LBFGS_1995} algorithms is used to minimize Eq.~\eqref{eq:total_energy_non_dim}. The network topology is updated using T1-moves\cite{Weaire_Foam_2001, Bi_SM_2014, bi_nphys_2015} during minimization. This combined algorithm produces final stable states  where the net residual force on vertices is less than $10^{-10}$. A wide range of parameters $\mu = 3.6-3.95$,  $\sigma = 0-0.3$ and $\kappa_A$ as well  a range of system sizes $N=36-900$ are simulated. 
%For energy-minimized states, shear modulus $G$ and edge tensions (Eq.~\ref{tension_def}) are calculated. 

\begin{figure}
\begin{center}
\includegraphics[width=1\columnwidth]{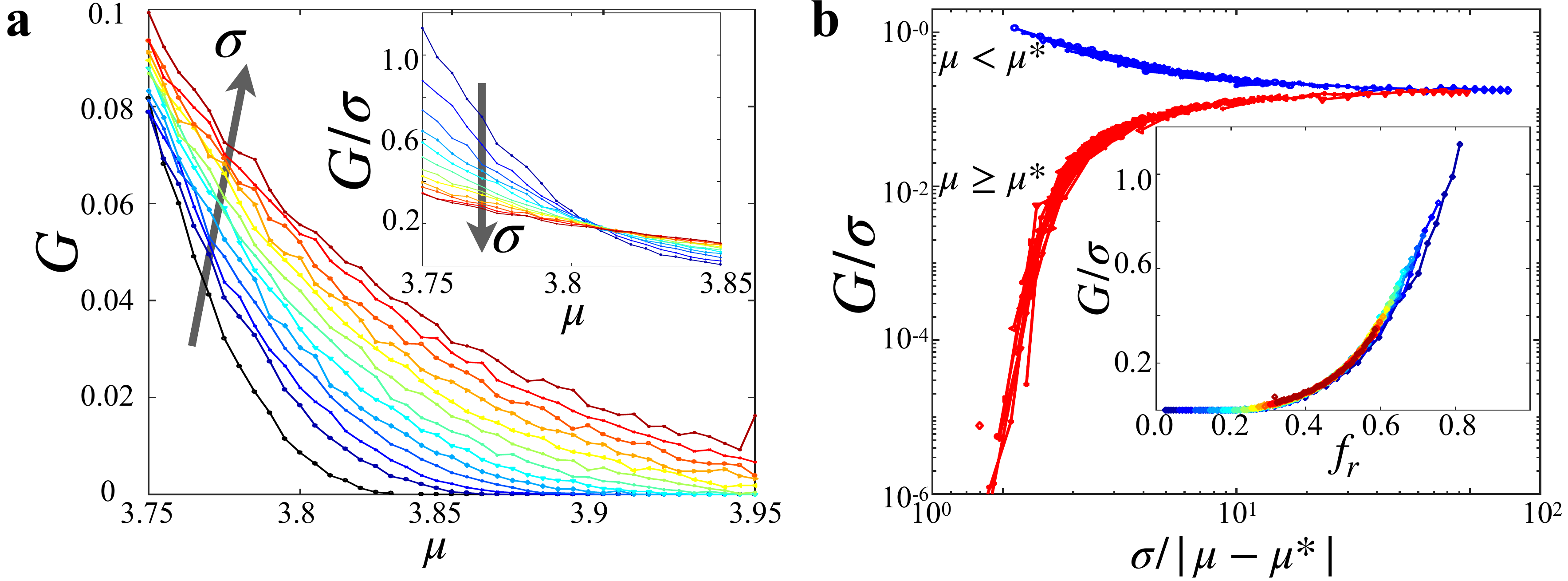}
 % AR = 2.70, 150/AR + 20 = 76 words
\caption{{\bf Tissue mechanical property.}
{\bf (a)} Shear modulus $G$ vs $\mu$ at various $\sigma=0,0.07,0.09,0.11,0.13,0.15,0.17,0.19,0.21,0.23,0.25,0.27,0.29$ for $N=400$ cells and $\kappa_A=1$.
{\bf Inset:} $G/\sigma$ vs $\mu$. 
{\bf (b)} $G/\sigma$ as function of $\sigma/|\mu-\mu^*|$.
{ \bf  Inset:}  $G/\sigma$ vs. the fraction of rigid cells $f_r$(Eq.~\ref{fr_def}). 
}
\label{fig1}
\end{center} 
\end{figure}

\begin{figure}[htp]
\begin{center}
\includegraphics[width=1\columnwidth]{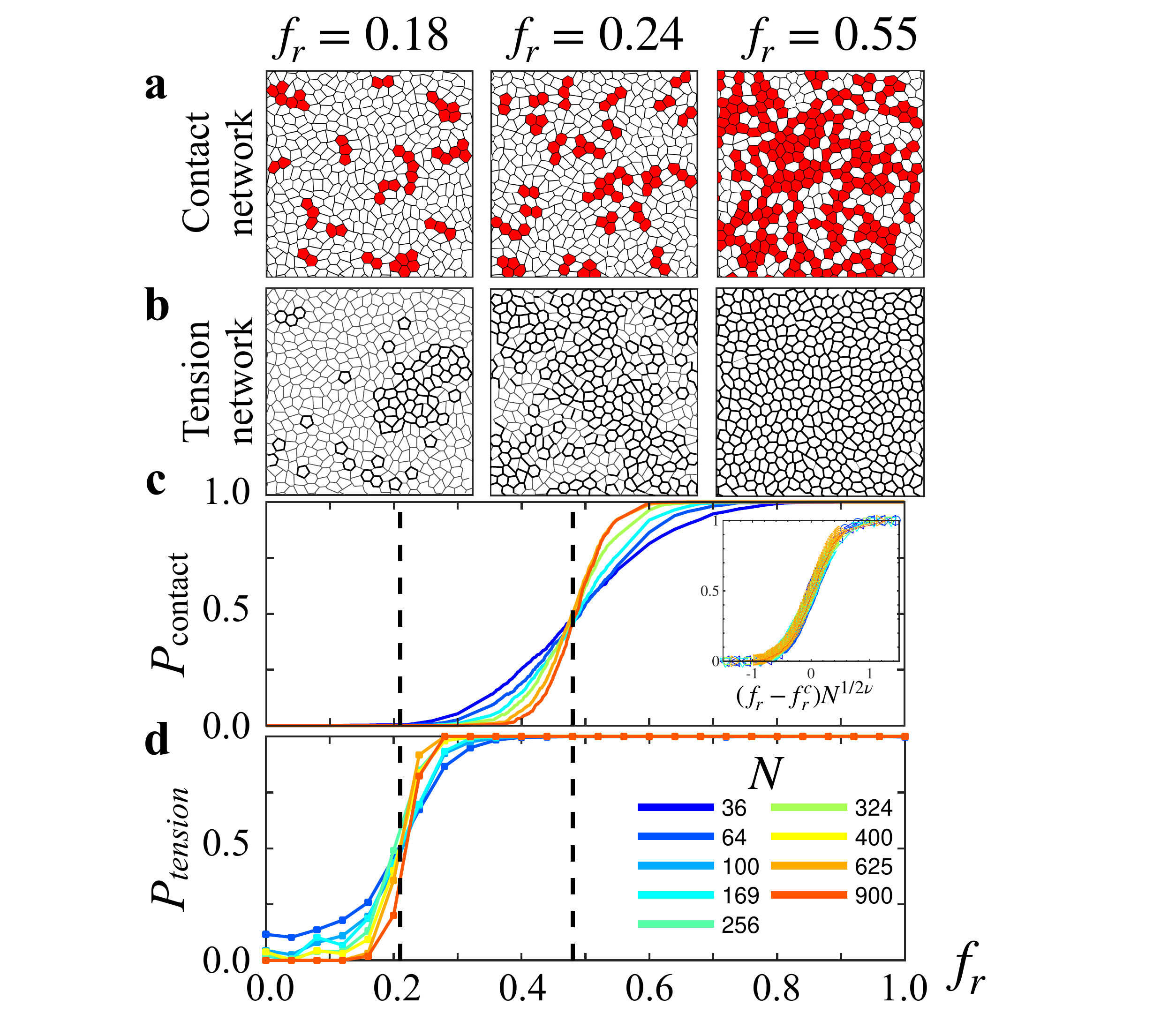}
% AR = 2.99, 150/AR + 20 = 71 words if single column
% AR = 2.99, 300/(2.99*0.5)+40=241 words if double column
% AR =1.15 150/AR + 20 = 150 words if single vertical column
\caption{
{\bf{Rigidity and contact percolations do not coincide.}}
{\bf (a)} Connected clusters of rigid cells shown in red. White cells correspond to $p_0^i\ge\mu^*$.
{\bf (b)} Typical snapshots for the tension network. Edges with finite tensions indicated by thick black lines while other edges have $\tau=0$.
{\bf (c)} Probability of contact percolation for rigid cells,$P_{\text{contact}}$  vs. $f_r$. Colors indicate different tissues sizes ranging from $N=36$ to $900$ (methods for calculating  $P_{\text{contact}}$ are detailed in SI Section II).  
Inset: finite-size scaling yields a contact percolation threshold of $f_{r}^{c} = 0.48 \pm 0.004$ and $\nu=1.62 \pm 0.01$.
{\bf (d)} Probability to obtain a system-spanning tension cluster vs. $f_r$. Finite-size analysis yields a transition at $f_{r}^{*} = 0.21 \pm 0.01$. 
}
\label{fig2}
\end{center}
\end{figure}

{\bf{Cellular \hh enhances rigidity-}}
We first focus on the case where area and perimeter elasticities are matched in their strengths ($\kappa_A = 1$). In Fig.~\ref{fig1}(a),  the shear modulus $G$ is plotted as function of $\mu$ and $\sigma$. At $\sigma = 0$,  $G$ vanishes at $\mu \approx 3.828$. This recapitulates the rigidity transition in the absence of \hh \cite{Farhadifar_CB_2007, bi_nphys_2015, yan_bi_prx}. 
%However, with \hh,  the rigidity of the tissue can change drastically.  
Increasing $\sigma$ at fixed $\mu$  always enhances the rigidity, which can occur in two different ways: (i) for a rigid tissue when $\sigma = 0$, $G$ always increases with $\sigma$, and (ii) for a fluid tissue, $G=0$ when $ 0< \sigma \le \sigma^*$  but becomes rigid when $\sigma \ge \sigma^*$. The threshold $\sigma^*$ where rigidity emerges depends on $\mu$. To better understand the dependence of $G$ on  $(\mu, \sigma)$, we first hypothesized that $\sigma$ provides an overall scale for the shear modulus and plot a rescaled $G/\sigma$ as function of $\mu$ for various $\sigma$ (Fig.~\ref{fig1}(a, inset)). This yields an intriguing result, where all curves in Fig.~\ref{fig1}(a, inset) intersect at a common point $\mu^* \approx 3.8$. 
The point $\mu^*$ serves as a cross-over point between two distinct  regimes. 
When $\mu < \mu^*$, $G/\sigma$ decreases as  $\sigma$ is increased.  
For $\mu \ge \mu^*$ the behavior is flipped and $G/\sigma$ \emph{increases} with $\sigma$. The intersection of all curves also suggests that closer to $\mu^*$, the shear modulus should converge to a scaling of $G \propto \sigma$. Based on these observations, we hypothesized that the behavior of $G$ below and above $\mu^*$ may be described by a universal scaling relation 
\be
G = \sigma \ g_{\pm} \left(\frac{\sigma}{\vert \mu-\mu^*\vert} \right).
\label{g_scaling}
\ee
We re-plot all data based on the ansatz (Eq.~\ref{g_scaling}) in Fig.~\ref{fig1}(b). An excellent scaling collapse is obtained, which also allowed us to pinpoint the location of the crossover transition $\mu^*=3.812$. The two distinct branches ($+$, $-$) of the scaling function $g$ correspond to  $\mu > \mu^*$ and $\mu < \mu^*$, respectively. The two branches  meet at $\mu = \mu^*$, where $G$ scales linearly with $\sigma$. 

What does this universal scaling in the vicinity of a cross-over transition point $\mu^*$ inform us about the nature of rigidity? 
At the crossover $\mu = \mu^*$ , there are  $50\%$ of cells with $p_0^i<\mu^*$ regardless of the value of $\sigma$. 
Below the crossover ($\mu < \mu^*$), cells with $p_0^i < \mu^*$ exceed $50\%$, while above the crossover ($\mu > \mu^*$) this fraction is below $50\%$. This suggests that the fraction of cells with $p_0^i < \mu^*$ plays an important role. Therefore we define this fraction cells as
\be
f_r = \int_{-\infty}^{\mu^*} \mathcal{F}_{\mu,\sigma}(p_0) d {p_0}, 
\label{fr_def}
\ee
where $\mathcal{F}$ is the probability distribution function and for a Gaussian, 
$f_r= (1/2) \text{erfc} [ (\mu-\mu^*)/(\sqrt{2}\sigma)]$. In Fig.~\ref{fig1}(b-inset), a re-plot all data in terms of $G/\sigma$ vs. $f_r$  collapses to a common curve.
Interestingly, the numerical value $p_0 = 3.812$ was previously found to be the threshold for rigidity in  tissues\cite{bi_nphys_2015} with homogeneous $p_0$. 
This universal scaling behavior supports the idea that $p_0^i<3.812$ can serve as a \emph{single-cell} criterion for rigidity and their  fraction $f_r$ determines tissue mechanics. We therefore refer to $f_r$ as the \emph{fraction of rigid cells}. 

\begin{figure}[htbp]
\begin{center}
\includegraphics[width=1\columnwidth]{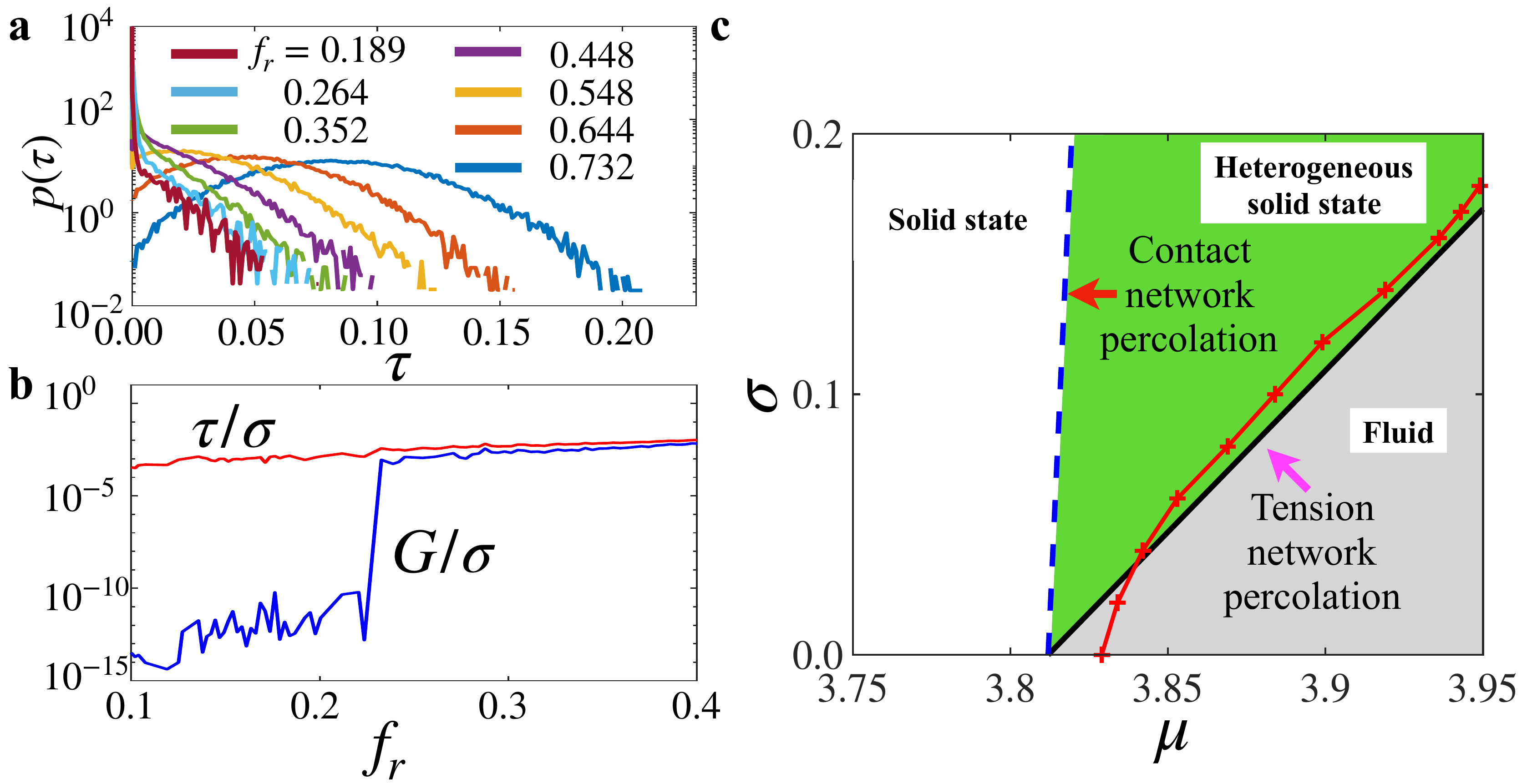}
\caption{
% AR = 1.95, 150/AR + 20 = 97 words
{\bf Spatial distribution of cellular tensions dictates tissue rigidity.}
{\bf   (a)} Distribution of tension at $\kappa_A=1$, $N=400$ for various $f_r$. 
{\bf   (b)}  $G/\sigma$ and  $\tau/\sigma$ versus $f_r$ at $\kappa_A=1$, $\sigma=0.1$, $N=400$. $G/\sigma$ vanishes at $f_r^*=0.21$ while the tension remains finite even in fluid states.
{\bf    (c)} A  phase diagram in $(\mu,\sigma)$. The solid black line indicates rigidity percolation and corresponds to values of $\mu$ and $\sigma$ where $f_r^* =0.21$. The red-line is the meanfield estimate for the bond-percolation on a random Voronoi tessellation. The blue dashed line indicates the contact percolation transition of rigid cells and corresponds to $f_r^c=0.48$. The mismatch between rigidity and contact percolation results in an intermediate phase or "heterogeneous solid state" which exists only for $\sigma>0$.  
}
\label{fig3}
\end{center}
\end{figure}

{\bf{Heterogeneity drives two different percolations-}}
 The emergence of rigidity when a certain population of cells exceeds a fraction is reminiscent of a percolation process\cite{Percolation_Pabitra_1984, Jacobs_PRL_1995, Thorpe_PRE_1996,Tang_PRB_1988,Ostojic_nature_2006}. To test if $f_r$ drives a percolation transition, we first  analyze the spatial organization of  rigid cells. Snapshots of connected rigid cells are shown in Fig.~\ref{fig2}(a) and the probability for rigid cells to form a system-spanning contact network is plotted in Fig.~\ref{fig2}(c). Finite-size scaling (Fig.~\ref{fig2}(c, inset)) shows a contact percolation transition at $f_r^c = 0.48$ with exponent $\nu = 1.62 \pm 0.01$.
 However, according to Fig.~\ref{fig1}(b,inset) and Fig.~\ref{fig3}(b), the tissue becomes rigid at a much lower $f_r\approx0.21$, suggesting that contact percolation of rigid cells is not necessary for rigidity.
 % 
% This leaves an intriguing question: how can rigidity emerge before the rigid cells percolate? 
We then hypothesized that the percolation of the tension network is the necessary condition for rigidity rather than rigid cell contacts. Fig.~\ref{fig2}(d) shows tension percolation occurring at $f_r^* = 0.21 \pm 0.01$, coinciding with the rigidity onset. Snapshots of tensions (Fig.~\ref{fig2}(b)) highlight an interesting behavior: e.g. at $f_r=0.24$, the rigid-cell contact network does not span the system, but the tensions form a system-spanning structure. Therefore the presence of just a few unconnected rigid cells can induce a much more spatially extended tension network. 
%
% This leaves an intriguing question: how can rigidity emerge before contact percolation can take place? 
%We next analyze the spatial distribution of tensions on cell edges. 
The distribution of $\tau$ also evolves as function of $f_r$ in an unconventional way (Fig.~\ref{fig3}(a)). 
%In  we show the distribution of $\tau$ as function of $f_r$. 
For values of $f_r >0.21$ corresponding to rigid states, $p(\tau)$ vanishes as $\tau\to0$ and appears symmetrical about their mean. At lower $f_r$, $p(\tau)$ becomes significantly more skewed and heavy-tailed (SI FIG.S2) and also develops an excess number of $\tau\approx0$ edges. Interestingly, the average tension for the tissue $\langle \tau \rangle$ does not vanish (Fig.~\ref{fig3}(b)) even below $f_r^*$, and does not show drastic changes at $f_r^*$. In contrast, $G/\sigma$ undergoes a transition $f_r^*$, coinciding with the tension network percolation. Remarkably, two states with slight differences in their tensions can differ by several orders of magnitudes in $G$(Fig.S3), depending on whether the tension network is percolated. 

Next we use a simple meanfield approach to estimate the rigidity transition threshold $f_r^* = 0.21$. Since our simulated tissues share a similar connectivity and topology with random Voronoi tessellations, we recall the \emph{bond percolation} threshold ($\sim 0.66$)  on such networks\cite{Vinod_AP_1971, Hsu_PRE_1999, Becker_PRE_2009}. Therefore if rigidity requires the percolation of edges with finite tensions,  the onset of rigidity should correspond to when there are exactly $66\%$ cell edges with finite tensions. For each parameter set $(\mu, \sigma)$ we define the fraction of finite tensions to be $N_\tau= \int_{\tau_0}^\infty p(\tau)d\tau$, where the threshold $\tau_0 = 10^{-5}$ is chosen to define ``zero" tensions (coinciding with the noise floor in the numerical calculations). In Fig.~\ref{fig3}(c) we show that the point at which $N_\tau(\mu,\sigma) = 0.66$ closely coincides with the rigidity transition, i.e. $f_r^* =0.21$. Taken together, these results suggest that a \emph{rigidity percolation} occurs at $f_r^*$. 

The mechanics of the tissue is summarized in a phase diagram (Fig.~\ref{fig3}(c)). Solid and fluid-like states are separated by $f_r^*$ which corresponds to a rigidity percolation of the edge tensions in the network. Contact percolation of rigid cells occur deeper within the solid phase at $f_r = 0.48$. 
The mismatch between contact and rigidity percolations can be observed only in the presence of \hh since at $\sigma=0$, $f_r$ is either $0$ or $1$.   We call the intermediate region between the two percolation transitions the {\emph{Heterogeneous solid state}}. Rigidity transition obtained through the meanfield method is shown  as the red line in Fig.~\ref{fig3}(c).

\begin{figure}[htbp]
\begin{center}
\includegraphics[width=1\columnwidth]{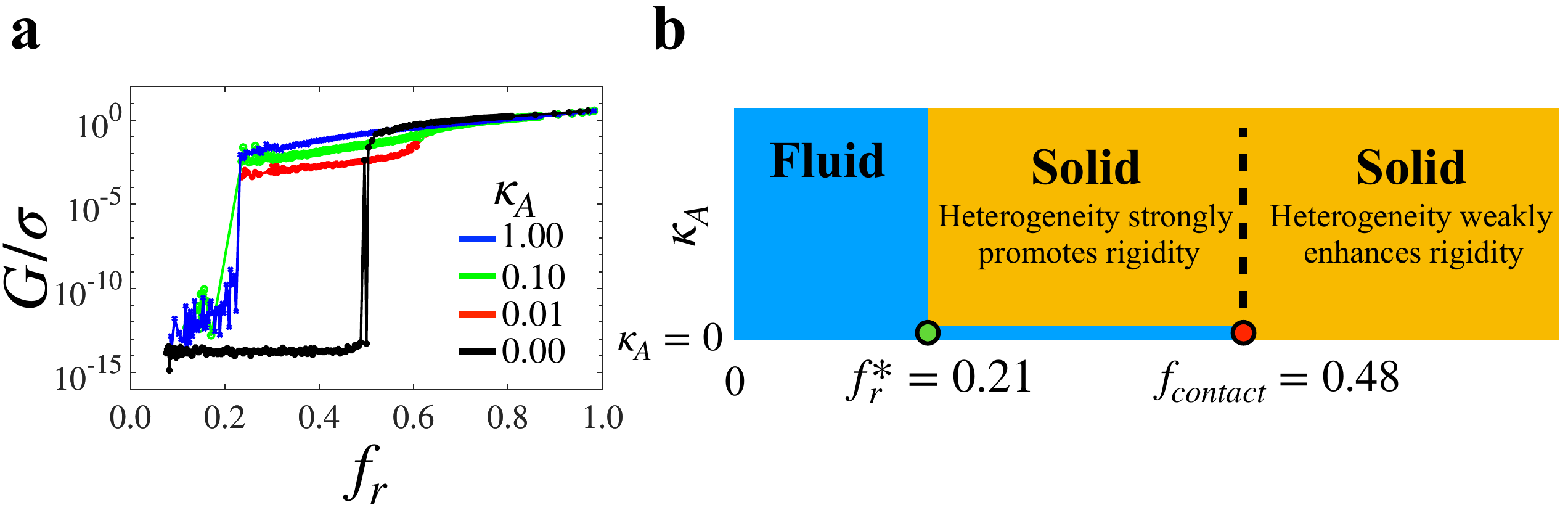}
% AR = 3.06, 150/AR + 20 = 69 words
\caption{
{\bf {The effect of \hh depends on cell-area elasticity.}}
{\bf   (a)}  $G/\sigma$ vs $f_r$ at various $\kappa_A$ values. 
{\bf   (b)} Phase diagram as function of $f_r$ and $\kappa_A$. The rigidity percolation  and contact percolations are distinct when $\kappa_A>0$, but coincide when  $\kappa_A=0$. The region of ($\kappa_A>0, 0.21<f_r<0.48$) corresponds to a solid that follows  the lower branch in Fig.~\ref{fig1}(b), where the shear modulus is sharply enhanced by increasing \hh. 
} 
\label{fig4}
\end{center}
\end{figure}

{\bf{Cell area elasticity $\kappa_A$ weakens the effect of heterogeneity-}}  
Next we focus on the {\emph{Heterogeneous solid state}} ($0.21 < f_r < 0.48)$. The regime is characterized by spatial heterogeneity (Fig.~\ref{fig2}(b)) where isolated cells or isolated clusters exist along side tissue-spanning percolated networks. Mechanical force balance must hold at every vertex\cite{yan_bi_prx} for a solid tissue. However,  for an isolated tension cluster, edge tensions alone cannot guarantee force balance, such as a vertex that sits on the boundary of this cluster. The missing component required for full force balance is the \emph{intracellular pressure} force that acts perpendicularly on an edge given by\cite{Yang_PNAS_2017,yan_bi_prx}  
$
\Pi_{m} =  \kappa_A (a_i-a_j). 
$
Here $a_{i}$ and $a_{j}$ are the areas of cell $i$ and $j$ adjacent to the edge $m$. Interestingly, since the pressure depends on the value of the cell area elasticity $\kappa_A$, the stability of isolated tension clusters must also depend on $\kappa_A$. To see the effect of cell area elasticity, we repeated the numerical calculations at various $\kappa_A$ values. The dependence of $G/\sigma$ on $f_r$ is plotted for various $\kappa_A$ in Fig.~\ref{fig4}(a). At $\kappa_A >0$, $G/\sigma$ suffers a slight dip at $f_r \sim 0.48$, but stays finite all the way until $f_r$ below $f_r^*$. In general, as long as $\kappa_A$ is finite, the rigidity transition always occurs at $f_r^*$ and does not shift. The case of $\kappa_A = 0$ is a singular limit where the rigidity transition shifts suddenly to $f_r=0.48$ to coincide with the contact percolation transition. This confirms our hypothesis that when cells cannot exert pressure forces, the tension network can only support mechanical rigidity when rigid cells \emph{physically come in contact}. In contrast, at $\kappa_A > 0$, stable tensions can be induced between rigid cells that are separated by a distance.  We summarize these results with in Fig.~\ref{fig4}(b) and also incorporate the different scaling regimes for the solid phase when $\kappa_A>0$. We are able to differentiate two types of solids: (i)when $\kappa_A>0,\ 0.21<f_r<0.48$, rigidity is strongly enhanced by heterogeneity (lower-branch in Fig.~\ref{fig1}(b)) and (ii) when $\kappa_A>0,\ f_r>0.48$, there is only weak enhancement of rigidity (upper-branch in Fig.~\ref{fig1}(b)). When $\kappa_A=0$, heterogeneity has no effect.

\begin{figure}[htbp]
\begin{center}
\includegraphics[width=1\columnwidth]{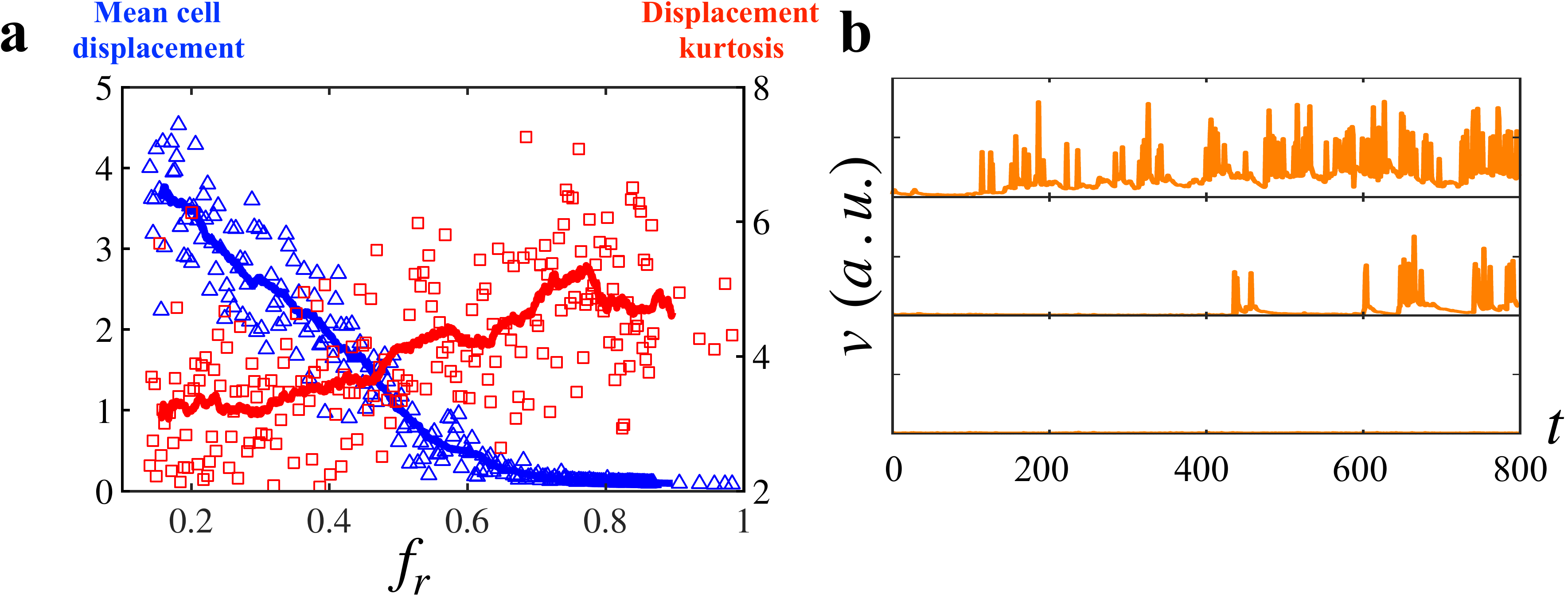}
% AR = 1.9, 150/AR + 20 = 78 words
\caption{
{\bf{Heterogeneity and cellular invasion}} 
{\bf    (a)} Analysis of the time-averaged displacement of  a single  cell attempting to move through the tissue with constant $v_0 = 0.4$ at various $f_r$.  Blue points represent the mean displacement of the cell taken over twice the persistence time. Red points represent the kurtosis of short time displacements indicating a growing intermittency of the invasion dynamics.
{\bf    (b) } Representative traces of the  cell displacements  for (i)the fluid phase($f_r = 0.14$), (ii)the intermediate solid phase($f_r=0.47$) and (iii)the rigid phase($f_r = 0.98$).
}
\label{fig5}
\end{center}
\end{figure}

{\bf Heterogeneity and cellular invasion -} 
Having established the static mechanical properties, we next focus on the effect of the heterotypic microenvironment on cell migration. We use a 
dynamic vertex model\cite{staddon_plos_2018,das_dvm_unpublished} to simulate the motion of a single invading cell in the tissue.   The invading cell  has a  propulsive force $v_0$ along a polarity vector $\bf{n}$, which undergoes random rotational diffusion\cite{fily_marche_prl_2012,Bi_PRX_2016} at a  slow rate. This mimics the directional motility of a metastatic cell under the influence of strong chemotactic signals\cite{clark_review_model_migration}. In the tissue each vertex $v$ evolves according to the overdamped equation of motion
\be
% 3 tows = 48 words
\frac{d\mathbf{r}_{\nu}}{dt} =
\begin{cases}
-\frac{\partial{\epsilon}}{\partial \mathbf{r}_\nu} + v_{0} \mathbf{n} & \text{vertices of invading cell},\\
-\frac{\partial{\epsilon}}{\partial \mathbf{r}_\nu} & \text{for other vertices.}
\end{cases}
\label{invasion}
\ee

Eq.~\ref{invasion} is simulated with a fixed value $v_0=0.4$ and at various values of $f_r$. The mean total displacement of the invading cell is plotted as function of $f_r$ in  Fig.~\ref{fig5}a, the behavior near $f_r=0$ and $f_r=1$  recapitulate results in the absence of \hh\cite{Bi_PRX_2016}, where a cell is either moving ($f_r=0$) or stuck ($f_r = 1$). However with \hh  tissues in the range of $0<f_r <1$ become accessible and cells moving through must interact with rigid as well as non-rigid neighbor cells along its path of invasion. This results in a highly intermittent migration dynamics for the invading cell. To quantify intermittency, we plot the kurtosis in the displacement of the invading cell\cite{das_prl_2016} as function of $f_r$. The displacement of the invading cells also show a burst-like dynamics when the tissue is in the heterogeneous solid state (Fig.~\ref{fig5}(b) and SI Video 1  \href{https://youtu.be/GTIcr0Wc-NQ}{https://youtu.be/GTIcr0Wc-NQ}). 

{\bf Discussion -}  

In cancer research it has long been recognized that tumors are stiffer than the surrounding tissue. Yet tumors often contain softer cells and softer cells facilitate invasion\cite{Oswald_2017}. This causes an apparent paradox: how can soft cells give rise to a tumor that is rigid at the tissue level? Our work suggests that tissue mechanics is not just determined by the average softness of the cells, but on the statistical fluctuations in single cell properties. The heterogeneity-driven rigidification  can provide a possible resolution of this paradox.  As a simple example, we note that in the phase diagram (Fig.~\ref{fig3}(c)), it is possible to transform from a fluid to a solid state by increasing $\mu$ (i.e. cells become softer overall) but at the same time increasing $\sigma$. Furthermore, our predictions show that the tissue mechanics is controlled by a single fraction $f_r$. This is consistent with recent experimental finding\cite{Wong_2016} that the fraction of mesenchymal cells can serve as a control parameter in describing jamming properties and that increasing mesenchymal fraction leads to the increase in motility. Finally, there is also experimental evidence that \hh can drive more intermittent dynamics in cell migration\cite{wirtz_pulsating_migration}. The burst-like intermittency in the heterogeneous solid state   is highly reminiscent of the pulsating cancer cell migration recently observed in epithelial monolayers\cite{wirtz_pulsating_migration}. 

\begin{acknowledgments}
We thank J. M. Schwarz, Xingbo Yang, Josef A K{\"a}s for insightful discussions. The authors acknowledge the support of the Northeastern University Discovery Cluster.  
\end{acknowledgments}
\bibliography{vm_heterogeneity.bib}

\end{document}